\documentstyle[12pt]{article}

\renewcommand{\baselinestretch}{1.5}

\def\beq{\begin{equation}}
\def\eeq{\end{equation}}
\def\bea{\begin{eqnarray}}
\def\nn{\nonumber \\ }
\def\eea{\end{eqnarray}}
\def\ds{\displaystyle}

\def\ms{\medskip}
\def\bs{\bigskip}
\def\ni{\noindent}

\def\req#1{(\ref{#1})}
\def\ie{{i.e.}\ }

\def\Tr{{\rm Tr}\ }

\def\slashed#1{\slash\!\!\!#1}

\begin{document}

\hfill UB-ECM-PF 94/2

\hfill February 1994

\vspace*{3mm}

\begin{center}

{\LARGE \bf
Schwinger-Dyson Equations and Chiral Symmetry Breaking in $2D$ Induced
Gravity}

\renewcommand\baselinestretch{0.8}

\vspace*{0.5cm}
{\rm {\sc E. Elizalde}\footnote{E-mail: eli@ebubecm1.bitnet,
eli@zeta.ecm.ub.es}
\\ Department ECM and IFAE, Faculty of Physics, University of Barcelona,
\\ Diagonal 647, 08028 Barcelona, \\
and Center for Advanced Studies, CSIC, Cam{\'\i} de Sta B{\`a}rbara, \\
17300 Blanes, Catalonia, Spain, \\
\vspace{0.2cm}
{\sc S.D. Odintsov}}\footnote{
On leave of absence from Dept. of Mathematics and Physics,
Pedagogical Institute, 634041 Tomsk, Russia.
E-mail: odintsov@ebubecm1.bitnet}
and {\sc A. Romeo} \\
Department ECM, Faculty of Physics, University of Barcelona, \\
Diagonal 647, 08028 Barcelona, Catalonia, Spain, \\
\vspace{0.2cm}
and \\
{\sc Yu.I. Shil'nov} \\
Department of Theoretical Physics, Kharkov State University, \\
Svobody sq. 4, Kharkov 310077, Ukraine.

\ms

\renewcommand\baselinestretch{1.4}

\vspace{5mm}

{\bf Abstract}

\end{center}

The Schwinger-Dyson equations in the ladder approximation for $2D$
induced gravity coupled to fermions on a flat background are obtained in
conformal gauge. A numerical study of these equations shows the possiblity of
chiral symmetry breaking in this theory.

\newpage

The Schwinger-Dyson equations provide a convenient way to study
dynamical symmetry breaking and dynamical mass generation in quantum
field theory. However, they constitute an infinite set of integral
equations and surely some truncation scheme is necessary in order to be
able to solve it.

In pioneering works \cite{Mas,Fuk}, the approach to dynamical
fermion
mass generation in quantum electrodynamics based on some truncated
version of the Schwinger-Dyson equations (sometimes called
ladder approximation) was developed, and it was shown the possibility
of chiral symmetry breaking in QED, and also the existence of a critical
coupling
constant in the Landau gauge (for a review see \cite{Fom}). During the
last fifteen years there has been much activity in developing and
further extending the approach of refs. \cite{Mas,Fuk} to dynamical symmetry
breaking in QED (for a review of the current status of this subject, see
the proceedings \cite{Mut}).
However,
if dynamical mass is generated,
the nonperturbative Ward-Takahashi identity in QED is not satisfied.
That is
why the dynamically generated fermionic mass, as well as the critical
coupling constant in QED, are highly gauge-dependent. That fact was
confirmed recently by the study of dynamical symmetry breaking in QED in
an arbitrary covariant gauge \cite{Aok}.

Currently, there is no doubt that the Schwinger-Dyson equations are
powerful
tools for the study of non-perturbative effects in field theory.
However,
perhaps some new approaches to truncation of the Schwinger-Dyson
equations
should be developed, particularly in order to solve the gauge-dependence
problem in the dynamical symmetry breaking of QED. It may also be useful
for this purpose to study the standard Schwinger-Dyson equations
\cite{Mas,Fuk} in other models (albeit more complicated) like quantum
gravity. Some time ago, a study along this line was done
in Ref. \cite{Abe} for $4D$
Einstein quantum gravity coupled to fermions on a flat background. As a
result, the possibility of chiral symmetry breaking in
such model was shown \cite{Abe}.

The purpose of the present letter is to study the Schwinger-Dyson
equations and the problem of chiral symmetry breaking in $2D$ induced
gravity \cite{Pol} coupled to fermions on a flat background. Notice
that,
recently, $2D$ induced gravity has been a popular subject of study as a
toy model for more realistic $4D$ gravity, which still does not exist as
a self-consistent theory.

We shall start from the following action
\bea
S&=&S_g+S_f, \nn
S_g&=&\ds -{1 \over 2 \gamma} \int d^2x \ \sqrt{-g}
\left( R {1 \over \Delta} R + \Lambda \right) , \nn
S_f&=&\ds \int d^2x \ \sqrt{-g} \ i \ \bar\Psi \gamma^{\mu} D_{\mu} \Psi,
\label{act}\eea
where $R$ is the two-dimensional curvature, $\Psi$ the $2D$ spinor,
$\Lambda$ the cosmological constant, and $D_{\mu}$ is the $2D$
covariant derivative for spinors. In the standard approach to $2D$
induced gravity, working in the conformal gauge \cite{Dav,Dis}
\beq g_{\mu \nu}=e^{\varphi}\eta_{\mu \nu}, \label{cfg}\eeq
one can start from a matter theory (like $S_f$) and induce $S_g$ from
it, by integrating over matter. After that, one has an exactly solvable
theory with the action $S_g$.

Instead, we will adopt here another approach, namely, to start from the
action \req{act},
where gravitational and spinor fields are supposed to be quantum. Using
the conformal gauge \req{cfg} and working in the {\it vierbein}
formalism
on a flat background, we consider $\gamma$ as some given constant and do
not integrate over $\Psi$.
Then, the term of the action corresponding to the interaction between
$\Psi$ and $\varphi$ is given by \beq
S_{\rm int}={i \over 2}\int d^2x \ \bar\Psi
\left[ \varphi \slashed{\partial} \Psi
+ {1 \over 2} ( \slashed{\partial} \varphi ) \Psi \right].
\label{Sint}\eeq

Let us now calculate the effective potential for composite fields
\cite{Cor} in the ladder approximation \cite{Mas}-\cite{Fom}:
\beq
V_{\rm eff}= -i \ {\rm Sp} [ \ln S_0^{-1}S - S_0^{-1} S +1 ] + V_2 ,
\label{Veff1}\eeq
where the free fermion propagator is given by
\beq S_0(p)={1 \over \slashed{p}}, \label{S0} \eeq
and the exact fermion propagator is
\beq
S(p)={1 \over {\cal A}(p^2)\slashed{p} -{\cal B}(p^2)}, \label{Sfull}
\eeq
${\cal A}$ and ${\cal B}$ being, for the moment, unknown functions.

Note that $V_2$ corresponds (as it follows from the structure of
$S_{\rm int}$ \req{Sint}) to a two-particle irreducible diagram, which
is similar to the analogous diagram in QED. Here, in $V_2$ the fermion
propagator is the exact one, while the vertex and the $\varphi$
(graviton) field propagator
\bea
\Gamma(k,p)&=&\ds {1 \over 2}\slashed{p}+{1 \over 4}\slashed{k}, \nn
G(k)&=&\ds{ \gamma \over k^2-\Lambda/2 }.
\label{vertprop}\eea
are tree-level quantities (ladder approximation). Hence, $V_2$ is
given by \beq
V_2=-{i \over 2} \int
{d^2 p \over (4\pi)^2} \int {d^2 q \over (4\pi)^2}
\Tr \left[ \Gamma(p-q,q) S(q) \Gamma(q-p,p) G(p,q) \right] .
\label{V2ladder}
\eeq
Using \req{Sint} and \req{Sfull}-\req{V2ladder} for calculating the
effective potential \req{Veff1}, one can get, after performing Wick's
rotation and the angular integration (we drop the details of
these calculations):
\[
V_{\rm eff}=-{N_f M^2 \over 8 \pi}\left\{
\int_0^1 dx \left[ \ln\left( A^2(x) + { B^2(x) \over x } \right)
-2 { A(x)(A(x)-1)x+B^2(x) \over x A^2(x) + B^2(x) } \right] \right.
\]
\beq
\left. +g \int_0^1 {dx \over x A^2(x)+B^2(x) }
\int_0^1 {dy \over y A^2(y)+B^2(y) }
[ A(x)A(y)K_A(x,y)+B(x)B(y)K_B(x,y) ] \right\} ,
\label{VeffAB}\eeq
where $N_f$ is the dimension of the fermion representation, $M$ is the
momentum cutoff, $x=p^2/M^2$,
$y=q^2/M^2$, $A(x)={\cal A}(p^2)$, $B(x)={\cal B}(p^2)/M$  and
\bea
K_A(x,y)&=&\ds-{ 4xy+(x+y)(x+y+l-\sqrt{(x+y+l)^2-4xy}) \over
2\sqrt{(x+y+l)^2-4xy} }, \nn
K_B(x,y)&=&\ds{ 2(x+y)+l-\sqrt{(x+y+l)^2-4xy} \over
\sqrt{(x+y+l)^2-4xy} },
\eea
with the notations
\beq
g={\gamma \over 64 \pi}, \eeq  and \beq
l={\Lambda \over 2M^2 }.
\label{gl}\eeq

The Schwinger-Dyson equation which corresponds to the effective
potential of the composite fields
 \req{VeffAB}, in the ladder
approximation takes the form \beq
i( S^{-1}(p)-S_0^{-1}(p) )= \int {d^2 q \over (4\pi)^2}
\left[ \Gamma(p-q,q) S(q) \Gamma(q-p,p) G(p,q) \right] .
\label{SDeqSigma}
\eeq
Bearing in mind \req{Sfull}, integrating over the angles and doing
the same changes of variable as before in \req{SDeqSigma}, one can show
that the functions $A$ and $B$ must obey integral equations of the
following form
 \bea
A(x)&=&\ds 1+g\int_0^1 dy {A(y) \over yA^2(y)+B^2(y) }{1 \over x}
K_A(x,y), \nn
B(x)&=&\ds g\int_0^1 dy {B(y) \over yA^2(y)+B^2(y) } K_B(x,y) .
\label{SDAB}\eea
Of course, it is not possible to solve these equations analytically.
However, we will obtain their numerical solution by a standard iterative
method at some region of the theory parameters.
First, we fix the values of the parameters $g$ and $l$ appearing
in \req{gl} and prepare two types of trial functions (the procedure
is quite similar to the one used in ref. \cite{Abe}):
\[
\begin{array}{lll}
\mbox{(a)}&A^0(x)=c_1,&B^0(x)=0, \\
\mbox{(b)}&A^0(x)=c_1,&B^0(x)=c_2,
\end{array}
\]
where $c_1$ abd $c_2$ are constants between 0 and 1. The functions
$A^0(x)$
and $B^0(x)$ are the starting point of a self-consistent iterative
calculation in which one finds successive pairs of functions $A^i(x)$
and $B^i(x)$. To
be more precise, each pair is obtained from the previous
one by means of the recurrence
\bea
A^{i+1}(x)&=&\ds 1+g\int_0^1 dy {A^i(y) \over y{A^i}^2(y)+{B^i}^2(y) }
{1 \over x} K_A(x,y), \nn
B^{i+1}(x)&=&\ds g\int_0^1 dy {B^i(y) \over y{A^i}^2(y)+{B^i}^2(y) }
K_B(x,y) .
\label{NSDAB} \eea
The sequences formed by the $\{ A^i(x) \}$ and $\{ B^i(x) \}$ are
expected
to converge into the functions $A(x)$ and $B(x)$, respectively,
which are the desired solutions of \req{SDAB}. In practice, we judge the
degree of convergence of these series by the smallness of the squared
norms of the differences $A^{i+1}-A^i$ and $B^{i+1}-B^i$. In our
calculations, we have set bounds of orders $10^{-4}-10^{-6}$.
If for the given $g$ and $l$ there are solutions of both types, (a) and
(b), only the most stable of both by $V_{\rm eff}$ \req{VeffAB} must be
chosen as the one corresponding to the true vacuum.

We have executed this algorithm to solve \req{NSDAB}, starting from
trial functions (a) and (b) for fixed $l=0.5$ and varying $g$.
For very small $g$'s, both types lead to curves close to $A(x)=1$,
$B(x)=0$, \ie the chiral symmetric solution, as should be expected.
Moreover, their respective
potentials are practically undistinguishable. As $g$ increases,
the $V_{\rm eff}$
for the (a)-type solution appears to be just slightly higher than for
the (b)-type one, thus selecting the latter as the physical vacuum.
This happens until some specific value of $g$, around 0.1, is
reached, for which
the (a)-solution looks rather reluctant to converge. Before this
happens,
$V_{\rm eff}$ is marginally higher than for the (b)-solution. This seems
to indicate that the solution `lost' was not physical.

Typical curves representing the $A$ and $B$ functions obtained are
shown in Fig. 1. All of them correspond to (b)-solutions (\ie chiral
non-symmetric solutions) found for different values of $g$,
and for $l=0.5$. The starting constants where $c_1=c_2=0.5$, but
their precise value has no influence on the final form of $A$ and $B$.

As is clear, the value of $V_{\rm eff}$ corresponding to a solution
obtained for a given $g$ changes as we vary $g$.
$V_{\rm eff}$ vanishes as $g$ approaches zero, and increases
as $g$ grows. There is a specific $g$ for which it stops
growing, and shows a slow decrease for larger $g$'s. However, this
should be regarded cautiously, since the larger $g$ is, the closer we
are to regions where perturbative methods fail.

Hence, as we see from our numerical analysis, there is a possibility of
chiral symmetry breaking in $2D$ induced gravity coupled to fermions.
Our study has been done in the `physical' conformal gauge which, in some
sense, may be considered as the analogue of the `physical' Landau gauge
in QED.
The results of these numerical studies should be gauge dependent,
particularly because we do not pay attention to the nonperturbative
Ward-Takahashi identity resulting from general covariance. (Note that
this identity is not so easy to obtain explicitly in the noncovariant
conformal gauge). Hence, for a complete proof that chiral symmetry
breaking in
$2D$ induced gravity is a real physical phenomenon one has to repeat the
same study in some covariant gauge (of harmonic type \cite{Odi}, for
example). This is left to be done elsewhere.

\hskip2cm

\ni{\Large \bf Acknowledgements}

S.D.O. would like to thank T. Muta for helpful discussions.
This work has
been supported by CIRIT (Generalitat de Catalunya) and by DGICYT
(Spain), project no. PB90-0022.

\newpage

\ni{\Large \bf Figure Caption}
\bs

\ni{\bf Fig. 1}.

Plot of the functions $A$ and $B$  obtained as the (b)-type
solutions for
$g=$0.1, 0.2 and 0.25, keeping $l=0.5$ fixed. Notice how $B$ deviates
more and more from the $g=0$ solution ($B(x)=0$) as $g$ increases.
Although not shown in the figure, the curve keeps going up for larger
values of $g$.

\end{document}